\documentclass[a4paper,twocolumn,array]{revtex4}
\usepackage{graphicx}

\begin{document}

\title{Using strong isomorphisms to construct game strategy spaces}

\author{Michael J. Gagen}

\email{mjgagen@gmail.com}

\date{21 July 2012}

\begin{abstract}
When applied to the same game, probability theory and game
theory can disagree on calculated values of the Fisher
information, the log likelihood function, entropy gradients, the
rank and Jacobian of variable transforms, and even the
dimensionality and volume of the underlying probability
parameter spaces. These differences arise as probability theory
employs structure preserving isomorphic mappings when
constructing strategy spaces to analyze games.  In contrast,
game theory uses weaker mappings which change some of the
properties of the underlying probability distributions within
the mixed strategy space.  In this paper, we explore how using
strong isomorphic mappings to define game strategy spaces can
alter rational outcomes in simple games, and might resolve some
of the paradoxes of game theory.
\end{abstract}

\maketitle

\section{Introduction}

One possibly fruitful way to gain insight into the paradoxes of
game theory is to show that probability theory and game theory
analyze simple games differently.  It would be expected of
course that these two well developed fields should always
produce consistent results.  However, we will show in this paper
that probability theory and game theory can produce
contradictory results when applied to even simple games. These
differences arise as these two fields construct mixed and
behavioural strategy spaces differently.

The mixed strategy space of game theory is constructed,
according to von Neumann and Morgenstern \cite{vonNeumann_44},
by first making a listing of every possible combination of moves
that players might make and of all possible information states
that players might possess. This complete embodiment of
information then allows every move combination to be mapped into
a probability simplex whereby each player's mixed strategy
probability parameters belong to ``disjoint but exhaustive
alternatives, \dots subject to the [usual normalization]
conditions \dots and to no others." \cite{vonNeumann_44}.  The
resulting unconstrained mixed strategy space is then a
``complete set" of all possible probability distributions that
might describe the moves of a game
\cite{vonNeumann_44,Nash_50_48,Nash_51_28,Kuhn_1953,Hart_92_19}.
Further, the absence of any constraints other than for
normalization ensures ``trembles" or ``fluctuations" are always
present within the mixed strategy space so every possible pure
strategy probability distribution is played with non-zero (but
possibly infinitesimal) probability \cite{Selton_1975}.
Together, these properties of the mixed strategy space---a
complete set of ``contained" probability distributions, no
additional constraints, and ever present trembles---lead to
inconsistencies with probability theory.

In constructing a mixed strategy space, probability theory first
examines how subsidiary probability distributions can be
``contained" within a mixed space and whether the properties of
the probability distributions are altered as a result.
Probability theory uses isomorphisms to implement mappings of
one probability space into another space.  An isomorphism is a
structure preserving mapping from one space to another space. In
abstract algebra for instance, an isomorphism between vector
spaces is a bijective (one-to-one and onto) linear mapping
between the spaces with the implication that two vector spaces
are isomorphic if and only if their dimensionality is identical
\cite{Chatterjee_2005}.  When the preservation of structure is
exact, then calculations within either space must give identical
results. Conversely, if the degree of structure preservation is
less than exact, then differences can arise between calculations
performed in each space.  It is thus crucial to examine the
fidelity of the ``containment" mappings used to construct the
mixed spaces of game theory.

Probability theory defines isomorphic probability spaces as
follows.  First, a probability space ${\cal
P}=\{\Omega,\sigma,P\}$ consists of a set of events $\Omega$, a
sigma-algebra of all subsets of those events $\sigma$, and a
probability measure defined over the events $P$. Two probability
spaces ${\cal P}=\{\Omega,\sigma,P\}$ and ${\cal
P}'=\{\Omega',\sigma',P'\}$ are said to be {\em strictly
isomorphic} if there is a bijective map
$f:\Omega\rightarrow\Omega'$ which exactly preserves assigned
probabilities, so for all $e\in\Omega$ we have $P(e)=P'[f(e)]$.
A slight weakening of this definition defines an {\em
isomorphism} as a bijective mapping $f$ of some unit probability
subset of $\Omega$ onto a unit probability subset of $\Omega'$.
That is, the weakened mapping ignores null event subsets of zero
probability.  This definition and equivalent ones are given in
Refs. \cite{Ito_1984,Gray_2009,Walters_1982}. In particular, we
note that strong isomorphisms between source and target
probability spaces require they have identical dimensionality
and tangent spaces \cite{Sernesi_1993}.

The mixed strategy space of game theory ``contains" different
probability distributions with many possessing different
dimensionality (according to probability theory).  Their altered
dimensionality within the mixed space can alter those computed
outcomes dependent on dimensionality.  A simple functional
illustration of this process can make this clear.  A
1-dimensional function $f(x)$ can be embedded within a
2-dimensional function $g(x,y)$ in two ways: using constraints
$g(x,y_0)=f(x)$, or limits $\lim_{y\rightarrow y_0}
g(x,y)=f(x)$. In either case, many of the properties of the
source function $f(x)$ are preserved, but not necessarily all of
them.  In particular, these different methods alter gradient
optimization calculations. That is, the gradient is properly
calculated when constraints are used, $f'(x)=g'(x,y_0)$, but not
when a limit process is used, $f'(x)\neq \lim_{y\rightarrow y_0}
\nabla g(x,y)$ (where $\nabla$ indicates a gradient operator).

In this paper, we will show that exactly the same discrepancies
arise when probability theory and game theory are applied to
simple probability spaces, and that these discrepancies can be
significant. It is useful to indicate the magnitude of these
discrepancies here to motivate the paper (with full details
given in later sections below).  We consider a simple card game
with two potentially correlated variables $x,y\in\{0,1\}$ with
joint probability distribution $P_{xy}$. In the case where $x$
and $y$ are perfectly correlated, probability theory (denoted by
P) and game theory (denoted by G) respectively assign different
dimensions to both the Fisher information matrix ($F$) and the
gradient of the log Likelihood function ($\nabla L$), and can
disagree on the value of the gradient of the joint entropy at
some points ($\nabla E_{xy}$):
\begin{equation}
  \begin{array}{c|cc}
                        & {\rm P}  &    {\rm G}    \\ \hline
  {\rm dim}(F)          &     1    &       3       \\
  {\rm dim}(\nabla L)   &     1    &       3       \\
   |\nabla E|           &     0    &     \infty. \\
  \end{array}
\end{equation}
These fields also disagree on the probability space gradients of
both the normalization condition ($P_{00}+P_{11}=1$) and the
requirement that the joint entropy equates to the marginal
entropy ($E_{xy}-E_x=0$):
\begin{equation}
  \begin{array}{c|cc}
                                      &     {\rm P}    &   {\rm G}       \\ \hline
  \nabla \left(P_{00}+P_{11}\right) &     0    &  \neq 0  \\
  \nabla \left(E_{xy}-E_x\right)    &     0    &  \neq 0. \\
  \end{array}
\end{equation}
Should these fields model a change of variable within this game,
they further disagree on the rank of the transform matrix ($A$),
and on the invertibility of the Jacobian matrix ($J$):
\begin{equation}
  \begin{array}{c|cc}
                  &     {\rm P}    &   {\rm G}       \\ \hline
  {\rm Rank}(A) &     1           &  2  \\
  J             &     {\rm Singular}    &  {\rm Invertible}. \\
  \end{array}
\end{equation}
These fields even disagree on the dimension ($d$) and volume
($V$) of the minimal probability space used to analyze the game:
\begin{equation}
  \begin{array}{c|cc}
      &  {\rm P}    &   {\rm G}       \\ \hline
  d &  1    &   3       \\
  V &  1    &   \frac{1}{6}.       \\
  \end{array}
\end{equation}
The differences between game theory and probability theory arise
due to the different use of isomorphic mappings to construct
mixed strategy spaces.

In Section
\ref{sect_Optimization_and_isomorphic_probability_spaces} we
show the necessity for considering isomorphic probability spaces
using examples ranging from simple dice games to bivariate
normal distributions. Section
\ref{sect_Mixed_and_behavioural_strategy_spaces} collects
results for the mixed and behavioural strategy spaces of a
simple two-stage game and again establishes the necessity for
taking account of isomorphic probability distributions.  We
apply these results in Section
\ref{sect_Optimizing_simple_decision_trees} to optimizing highly
nonlinear random functions over a decision tree involving
correlated variables. This section is then generalized and
applied to a strategic game in Section
\ref{sect_Optimizing_a_multistage_game_tree}. Throughout, we
place the details of many calculations within the Appendices to
show working and avoid cluttering the paper.

\begin{figure}[htb]
\centering
\includegraphics[width=0.8\columnwidth,clip]{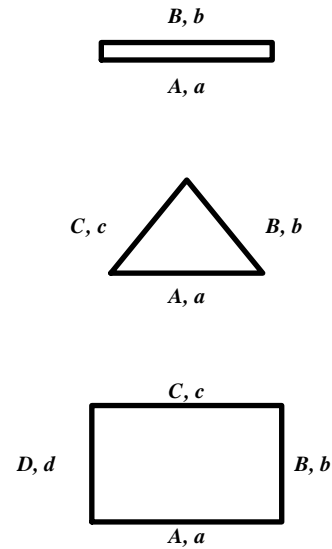}
\caption{\em Three alternate dice with different numbers of
sides. A coin with sides $A$ and $B$ appearing with respective
probabilities $a$ and $b$, a triangle with faces $A, B$ and $C$
occurring with respective probabilities $a, b$ and $c$, and a
square die with faces $A, B, C$ and $D$ each occurring with
respective probabilities $a, b, c$ and $d$.
 \label{f_alternate_dice}}
\end{figure}

\section{Optimization and isomorphic probability spaces}
\label{sect_Optimization_and_isomorphic_probability_spaces}

In this section, we introduce the need to use isomorphic
mappings when embedding probability spaces within mixed spaces.

\subsection{Isomorphic dice}

Consider the three alternate dice shown in Fig.
\ref{f_alternate_dice} representing a 2-sided coin, a 3-sided
triangle, and a 4-sided square.  Faces are labeled with capital
letters and the probabilities of each face appearing are labeled
with the corresponding small letter.  The corresponding
probability spaces defined by these die are
\begin{eqnarray}
  {\cal P}_{\rm coin}
    &=& \big\{ x\in\{A,B\},\{a,b\} \big\}  \nonumber \\
  {\cal P}_{\rm triangle}
    &=& \big\{ x\in\{A,B,C\},\{a,b,c\} \big\}  \nonumber \\
  {\cal P}_{\rm square}
    &=& \big\{ x\in\{A,B,C,D\},\{a,b,c,d\} \big\}.
\end{eqnarray}
Here the required sigma-algebras are not listed, and each of
these spaces are subject to the usual normalization conditions.
For notational convenience we sometimes write
$(p_1,p_2,p_3,p_4)=(a,b,c,d)$ and denote the number of sides of
each respective die as $n\in\{2,3,4\}$.

We now wish to optimize a nonlinear function over these spaces,
and we choose a function which cannot be optimized using
standard approaches in game theory. The chosen function is
\begin{equation}
  F = V^2 E_x,
\end{equation}
with
\begin{eqnarray}
  V  &=& \int_{\rm space} dv \nonumber \\
  E_x &=& - \sum_{i=1}^n p_i \log p_i,
\end{eqnarray}
where $V$ is the volume of each respective probability parameter
space and $E_x$ is the marginal entropy of each space
\cite{Georgii_2008}. We will complete this optimization in three
different ways, two of which will be consistent with each other
and inconsistent with the third.

As a first pass at optimizing the function $F$, we simply
maximize $F$ within each probability space and then compare the
optimal outcomes to determine the best achievable outcome.  As
is well understood, the entropy of a set of $n$ events is
maximized when those events are equiprobable giving a maximum
entropy of $E_{x, {\rm max}}=\log n$.  Using the volume results
of Eqs. \ref{eq_coin_functions}---\ref{eq_square_functions}, the
function $F$ takes maximum values in the three probability
spaces of
\begin{eqnarray}               \label{eq_F_Opt}
  F_{\rm coin,\; max} &=& \log 2 \nonumber \\
  F_{\rm triangle,\; max} &=& \frac{\log 3}{4} \nonumber \\
  F_{\rm square,\; max} &=& \frac{\log 4}{36}.
\end{eqnarray}
Comparing these outcomes makes it clear that the best that can
be achieved is to use a coin with equiprobable faces.

\begin{figure}[htb]
\centering
\includegraphics[width=0.9\columnwidth,clip]{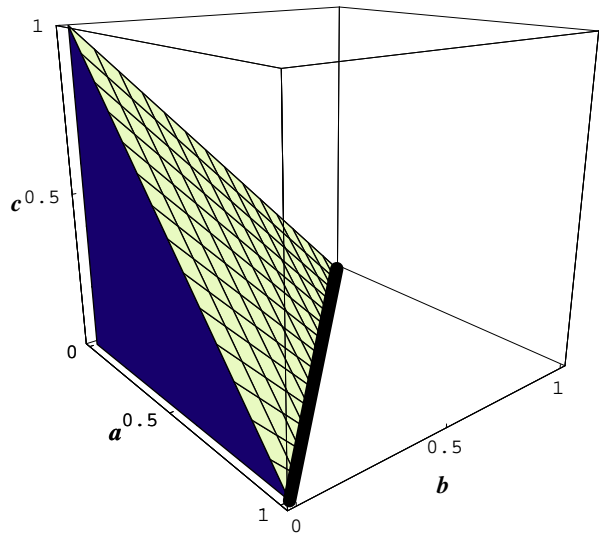}
\caption{\em The target space containing points corresponding to the
probability spaces respectively
of the coin ${\cal P}'_{\rm coin}$ along the line $a+b=1$ with constraint
$(c,d)=(0,0)$ (heavy line),
of the triangle ${\cal P}'_{\rm triangle}$ along the surface $a+b+c=1$
with constraint $d=0$ (hashed surface), and
of the square ${\cal P}'_{\rm square}$ filling the volume $a+b+c+d=1$
(filled polygon).
Note that points such as $(a,b,c)=(0.5,0.5,0)$ correspond to all three
probability spaces and are only distinguished by which constraints are acting.
 \label{f_mixed_dice}}
\end{figure}

The second method uses isomorphisms to map all of the three
incommensurate source spaces into a single target space.  We
choose our mappings as follows:
\begin{eqnarray}
  {\cal P}'_{\rm coin}
    &=& \left. \big\{ x\in\{A,B,C,D\},\{a,b,c,d\}
       \big\}\right|_{(cd)=(00)}  \nonumber \\
  {\cal P}'_{\rm triangle}
    &=& \left. \big\{ x\in\{A,B,C,D\},\{a,b,c,d\}
       \big\}\right|_{d=0}  \nonumber \\
  {\cal P}'_{\rm square}
    &=& \big\{ x\in\{A,B,C,D\},\{a,b,c,d\} \big\}.
\end{eqnarray}
Here, while all probability spaces share a common event set and
probability distribution, the isomorphic mappings impose
constraints on the ${\cal P}'_{\rm coin}$ and ${\cal P}'_{\rm
triangle}$ spaces.  The constraints arise from mapping the null
sets of zero probability from each source space to the
corresponding events of the enlarged target space.  The target
probability space is shown in Fig. \ref{f_mixed_dice} where the
normalization condition $d=1-a-b-c$ is used. The points
corresponding to the probability spaces of the coin ${\cal
P}'_{\rm coin}$ are mapped along the line $a+b=1$ with
constraint $(c,d)=(0,0)$. Those points corresponding to the
probability spaces of the triangle ${\cal P}'_{\rm triangle}$
are mapped along the surface $a+b+c=1$ with constraint $d=0$.
Finally, the probability spaces corresponding to the square
${\cal P}'_{\rm square}$ fill the volume $a+b+c+d=1$ and are not
subject to any other constraint.

The interesting point about the target space is that many
points, e.g. $(a,b,c,d)=(\frac{1}{2},\frac{1}{2},0,0)$, lie in
all of the probability spaces of the coin, triangle, and square
die and are only distinguished by which constraints are acting.
That is, when this point is subject to the constraint
$(c,d)=(0,0)$, then it corresponds to the probability space
${\cal P}'_{\rm coin}$ (and not to any other). Conversely, when
this same point is subject to an imposed constraint $d=0$ then
it corresponds to the probability space ${\cal P}'_{\rm
triangle}$. Finally, when no constraints apply then, and only
then does this point correspond to the probability space of the
square ${\cal P}'_{\rm square}$. This means that it is not the
probability values possessed by a point which determines its
corresponding probability space but the probability values in
combination with the constraints acting at that point.

It is now straightforward to use the isomorphically constrained
target space to maximize the function $F$ over all embedded
probability spaces using standard constrained optimization
techniques. For instance, to optimize $F$ over points
corresponding to the coin and subject to the constraint
$(c,d)=(0,0)$ then either simply resolve the constraint via
setting $c=d=0$ before the optimization begins, or simply
evaluate the gradient of $F$ at all points $(a,b,0,0)$ in the
direction of the unit vector $\frac{1}{\sqrt{2}}(-1,1,0,0)$
lying along the line $a+b=1$. (See Eq. \ref{eq_directed_grad}.)
An optimization over all three isomorphic constraints leads to
the same outcomes as obtained previously in Eq. \ref{eq_F_Opt}.
This completes the second optimization analysis and as promised,
it is consistent with the results of the first.

The same is not true of the third optimization approach which
produces results inconsistent with the first two.  The reason we
present this method is that it is in common use in game theory.
The third optimization method commences by noting that the
probability space of the square is complete in that it already
``contains" all of probability spaces of the triangle and of the
coin. This allows a square probability space to mimic a coin
probability space by simply taking the limit
$(c,d)\rightarrow(0,0)$.  Similarly, the square mimics the
triangle through the limit $d\rightarrow 0$. In turn, this means
that an optimization over the space of the square is effectively
an optimization over every choice of space within the square.
Specifically, game theory discards constraints to model the
choice between contained probability spaces. This optimization
over the points of the square has already been completed above.
When optimizing the function $F$ over the unconstrained points
corresponding to the square, the maximum value is $F=\log(4)/36$
at
$(a,b,c,d)=(\frac{1}{4},\frac{1}{4},\frac{1}{4},\frac{1}{4})$,
and according to game theory, this is the best outcome when
players have a choice between the coin, the triangle, or the
square.

The optimum result obtained by the third optimization method,
that used by game theory, conflicts with those found by the
previous two methods as commonly used in probability theory. The
difference arises as game theory models a choice between
probability spaces by making players uncertain about the values
of their probability parameters within any probability space.
Consequently, their probability parameters are always subject to
infinitesimal fluctuations, i.e. $c>0^+$ or $d>0^+$ always.
These fluctuations alter the dimensions of the space which
impacts on the calculation of the volume $V$ and alters the
calculated gradient of the entropy. Game theory eschews the role
of isomorphism constraints within probability spaces on the
grounds that any such constraints restrict player uncertainty
and hence their ability to choose between different probability
spaces. The probability parameter fluctuations mean that players
have access to all possible probability dimensions at all times
so a single mixed space is the appropriate way to model the
choice between contained probability spaces. In contrast,
probability theory holds that the choice between probability
spaces introduces player uncertainty about which space to use,
but specifically does not introduce uncertainty into the
parameters within any individual probability space.  As a
result, probability theory employs isomorphic constraints to
ensure that the properties of each embedded probability space
within the mixed space are unchanged.

The upshot is that a game theorist cannot evaluate the Entropy
(or uncertainty) gradient of a coin toss while considering
alternate die because uncertainty about which dice is used
bleeds into the Entropy calculation. However, the probability
theorist will distinguish between their uncertainty about which
face of the coin will appear and their uncertainty about which
dice is being used.

\subsection{Continuous bivariate Normal spaces}

The above results are general.  When source probability spaces
are embedded within target probability spaces, then the use of
isomorphic mapping constraints will preserve all properties of
the embedded spaces.  Conversely, when constraints are not used
then some of the properties of the embedded spaces will not be
preserved in general. We illustrate this now using normally
distributed continuous random variables.

Consider two normally distributed continuous independent random
variables $x$ and $y$ with $x,y\in(-\infty,\infty)$. When
independent, these variables have a joint probability
distribution $P_{xy}$ which is continuous and differentiable in
six variables, $P_{xy}(x,\mu_x,\sigma_x,y,\mu_y,\sigma_y)$ where
the respective means are $\mu_x$ and $\mu_y$ and the variances
are $\sigma_x^2$ and $\sigma_y^2$.  The marginal distributions
are $P_{x}(x,\mu_x,\sigma_x)$ and $P_{y}(y,\mu_y,\sigma_y)$.
(See Eq. \ref{eq_bivariate_normal_ind}.)

The independent joint distribution $P_{xy}$ can now be embedded
into an enlarged distribution representing two potentially
correlated normally distributed variables $x$ and $y$.  This
enlarged distribution
$P'_{xy}(x,\mu_x,\sigma_x,y,\mu_y,\sigma_y,\rho)$ differs from
$P_{xy}$ in its dependence on the correlation parameter
$\rho_{xy}=\rho$ with $\rho\in(-1,1)$.  This distribution is
continuous and differentiable in seven variables. (See Eq.
\ref{eq_bivariate_normal_corr}.) An isomorphic embedding
requires that the unit probability subset of $P_{xy}$ be mapped
onto the unit probability subset of $P'_{xy}$ and this is
achieved by imposing an external constraint that $\rho=0$ in the
enlarged space.  Hence, we expect
$\left.P'_{xy}\right|_{\rho=0}=P_{xy}$. It is readily confirmed
that when the isomorphism constraint is imposed on the enlarged
distribution all properties are preserved, while this is not the
case in the absence of the constraint.  The probability
distributions must satisfy a number of gradient relations (with
the gradient operator $\nabla$ a function of seven variables),
for instance
\begin{eqnarray}        \label{eq_bivariate_normal_corr_test}
   \left. \nabla \left[P'_{xy}-P'_xP'_y\right]\right|_{\rho=0}
         &=& 0 \nonumber \\
   \lim_{\rho\rightarrow 0} \nabla \left[P'_{xy}-P'_xP'_y\right]
     &\neq&  0  \nonumber \\
   \left. \nabla\left[P'_{x|y}-P'_x\right]\right|_{\rho=0}  &=& 0 \nonumber \\
   \lim_{\rho\rightarrow 0} \nabla\left[P'_{x|y}-P'_x\right]
      &\neq&   0.
\end{eqnarray}
(See Eq. \ref{eq_bivariate_normal_corr_t2}.) Similarly, the
expectations of functions of the $x$ and $y$ variables must also
satisfy a number of gradient relations (with the gradient
operator $\nabla$ now a function of five variables), for
instance
\begin{eqnarray}        \label{eq_bivariate_normal_corr_test2}
   \left. \nabla\left[ \langle xy\rangle'-\langle x\rangle' \langle y\rangle'\right]\right|_{\rho=0}
        &=& 0  \nonumber \\
   \lim_{\rho\rightarrow 0} \nabla\left[ \langle xy\rangle'-\langle x\rangle' \langle y\rangle'\right]
       &\neq& 0.
\end{eqnarray}
(See Eq. \ref{eq_bivariate_normal_corr_t3}.)

\begin{figure}[htb]
\centering
\includegraphics[width=0.8\columnwidth,clip]{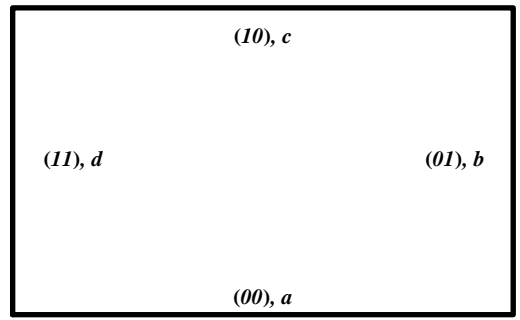}
\caption{\em A four-sided square probability space where joint variables
$x$ and $y$ take values $(x,y)\in\{(0,0),(0,1),(1,0),(1,1)\}$ with respective
probabilities $(a,b,c,d)$.
 \label{f_square_joint_xy_var}}
\end{figure}

\subsection{Joint probability space optimization}

We will briefly now examine isomorphisms between the joint
probability spaces of two arbitrarily correlated random
variables.  In particular, we consider two random variables
$x,y$ as appear on the square dice of Fig.
\ref{f_square_joint_xy_var} with probability space
\begin{eqnarray}
  {\cal P}_{\rm square}
    &=& \big\{ (x,y)\in\{(0,0),(0,1),(1,0),(1,1)\}, \nonumber \\
    && \hspace{1cm}  \{a,b,c,d\} \big\}.
\end{eqnarray}
The correlation between the $x$ and $y$ variables is
\begin{eqnarray}
 \rho_{xy}
    &=& \frac{\langle xy\rangle-\langle x\rangle\langle y\rangle}{\sigma_x \sigma_y} \nonumber \\
    &=& \frac{ad-bc}{\sqrt{(c+d)(a+b)(b+d)(a+c)}}.
\end{eqnarray}
Here, $\sigma_x$ and $\sigma_y$ are the respective standard
deviations of the $x$ and $y$ variables.

The space ${\cal P}_{\rm square}$ of course contains many
embedded or contained spaces.  We will separately consider the
case where $x$ and $y$ are perfectly correlated, and where they
are independent. As noted previously, there are two distinct
ways for these spaces to be contained within ${\cal P}_{\rm
square}$, namely using isomorphism constraints or using limit
processes.  These two ways give the respective definitions for
the perfectly correlated case
\begin{eqnarray}
  {\cal P}_{\rm corr}
    &=& \big\{ (x,y)\in\{(0,0),(0,1),(1,0),(1,1)\}, \nonumber \\
    && \left. \hspace{1cm}
             \{a,b,c,d\} \big\}\right|_{b=c=0}  \nonumber \\
  {\cal P}'_{\rm corr}
    &=& \lim_{(bc)\rightarrow(00)} \big\{ (x,y)\in\{(0,0),(0,1),(1,0),(1,1)\}, \nonumber \\
    && \hspace{1cm}   \{a,b,c,d\} \big\}
\end{eqnarray}
and for the independent case
\begin{eqnarray}
  {\cal P}_{\rm ind}
    &=& \big\{ (x,y)\in\{(0,0),(0,1),(1,0),(1,1)\}, \nonumber \\
    && \left. \hspace{1cm}
             \{a,b,c,d\} \big\}\right|_{ad=bc} \nonumber \\
  {\cal P}'_{\rm ind}
    &=& \lim_{ad\rightarrow bc} \big\{ (x,y)\in\{(0,0),(0,1),(1,0),(1,1)\}, \nonumber \\
    && \hspace{1cm}  \{a,b,c,d\} \big\}.
\end{eqnarray}
Here, all spaces satisfy the normalization constraint
$a+b+c+d=1$, which we typically resolve using $d=1-a-b-c$.
Evaluating any function dependent on a gradient or completing an
optimization task using either isomorphic constraints or limit
processes can naturally result in different outcomes as we now
illustrate.

\subsubsection{Perfectly correlated probability spaces}

We first consider the case where the $x$ and $y$ variables are
perfectly correlated in the spaces ${\cal P}_{\rm corr}$ with
isomorphism constraints or ${\cal P}'_{\rm corr}$ using limit
processes.

The maximum achievable joint entropy \cite{Georgii_2008} for our
two perfectly correlated variables obviously occurs at the point
where they are equiprobable.  This can be found by evaluating
the gradient of the joint entropy function
\begin{equation}
  E_{xy}(a,b,c) = -\sum_{xy} P_{xy} \log P_{xy}.
\end{equation}
In the space ${\cal P}_{\rm corr}$, the gradient optimization
$\nabla E_{xy}|_{b=c=0}=0$ locates an optimum point at
$(a,b,c,d)=(\frac{1}{2},0,0,\frac{1}{2})$, while in the space
${\cal P}'_{\rm corr}$ the optimum at $\nabla E_{xy}=0$ locates
the point
$(a,b,c,d)=(\frac{1}{4},\frac{1}{4},\frac{1}{4},\frac{1}{4})$.
(See Eq. \ref{eq_entropy_max}.)

The Fisher Information is defined in terms of probability space
gradients as the amount of information obtained about a
probability parameter from observing any event
\cite{Georgii_2008}. It is a matrix $F_{ij}$ with elements
$i,j\in\{1,2,3\}$. In the isomorphically constrained space
${\cal P}_{\rm corr}$, the Fisher Information is a scalar via
\begin{equation}
  F_{ij}|_{b=c=0} = F_{11} =  \frac{1}{a(1-a)},
\end{equation}
equal to the inverse of the Variance as required.  A very
different result is obtained in the unconstrained space ${\cal
P}'_{\rm corr}$ where the Fisher Information is a much larger
matrix. (See Eq. \ref{eq_Fisher_3d}.)

Probability parameter gradients also allow estimation of
probability parameters by locating points where the Log
Likelihood function is maximized $\nabla \log L=0$
\cite{Georgii_2008}. This evaluation takes very different forms
in the isomorphically constrained space ${\cal P}_{\rm corr}$
and the unconstrained space ${\cal P}'_{\rm corr}$ as shown in
Eq. \ref{eq_max_likelihood_3d}. Coincidentally however, in our
case the same estimated outcomes can be achieved in both spaces.
For example, if an observation of $n$ trials shows $n_a$
instances of $(x,y)=(0,0)$ and $n-n_a$ instances of
$(x,y)=(1,1)$ then both constrained and unconstrained approaches
give the best estimates of the probability parameters of
$(a,b,c,d)=(\frac{n_a}{n},0,0,1-\frac{n_a}{n})$.

Finally, when $x$ and $y$ are perfectly correlated it is
necessarily the case that expectations satisfy $\langle
x\rangle-\langle y\rangle=0$, that variances satisfy
$V(x)-V(y)=0$, that the joint entropy is equal to the entropy of
each variable so $E_{xy}-E_x=0$, and that finally, the
correlation between these variables satisfies $\rho_{xy}-1=0$.
All of these properties lead to gradient relations in the ${\cal
P}_{\rm corr}$ and ${\cal P}'_{\rm corr}$ spaces of:
\begin{eqnarray}
 \nabla\left[\langle x\rangle-\langle y\rangle\right]|_{b=c=0} &=& 0 \nonumber \\
  \lim_{(bc)\rightarrow (00)}\nabla\left[\langle x\rangle-\langle y\rangle\right] &=&
       -\hat{b}+\hat{c}  \nonumber \\
 \nabla\left[V(x)-V(y)\right]|_{b=c=0} &=& 0 \nonumber \\
 \lim_{(bc)\rightarrow (00)}\nabla\left[V(x)-V(y)\right] &=&
      (1-2a)\hat{b}-(1-2a)\hat{c} \nonumber \\
 \nabla\left[E_{xy}-E_{x}\right]|_{b=c=0} &=& 0 \nonumber \\
 \lim_{(bc)\rightarrow (00)}\nabla\left[E_{xy}-E_{x}\right] &\neq &
      {\rm undefined} \nonumber \\
 \nabla \rho_{xy}|_{b=c=0} &=& 0 \nonumber \\
 \nabla \rho_{xy} &\neq & 0.
\end{eqnarray}
Obviously, taking the limit $(b,c)\rightarrow (0,0)$ does not
reduce the limit equations to the required relations.  (See Eq.
\ref{eq_expectation_relations_3d}.)

\begin{figure}[htbp]
\centering
\includegraphics[width=0.8\columnwidth,clip]{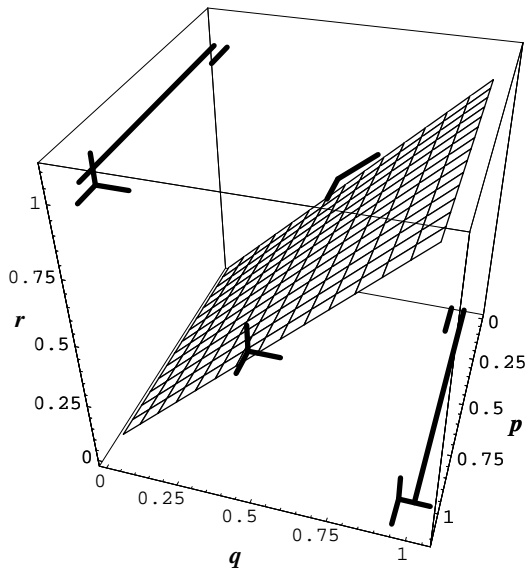}
\caption{\em  A schematic representation where a three dimensional
target probability strategy space $(p,q,r)$ embeds
respectively several one dimensional probability spaces associated
with perfectly correlated variables (lines, upper left and lower
right), and a two dimensional probability space associated with
independent variables (plane, middle). An exact isomorphism preserves
the respective original tangent spaces shown via one and two dimensional
axes offset in background.  A weak isomorphism fails to preserve the original
tangent spaces of the source probability distributions and assigns the
three dimensional tangent space of the target space to every embedded
distribution (as shown in foreground slightly
offset from each embedded space).
 \label{f_tangent_spaces}}
\end{figure}

\subsubsection{Independent probability spaces}

We next consider the case where the $x$ and $y$ variables are
independent using the spaces ${\cal P}_{\rm ind}$ with
isomorphism constraints or ${\cal P}'_{\rm ind}$ with limit
processes.

When random variables are independent, then their joint
probability distribution is separable for every allowable
probability parameter of ${\cal P}_{\rm ind}$ or ${\cal P}'_{\rm
ind}$.  This means the gradient of this separability property
must be invariant across both probability spaces. That is, we
must have both $P_{xy}=P_xP_y$ everywhere and hence
$\nabla\left[P_{xy}-P_xP_y\right]=0$.  Similarly, separability
requires we also satisfy $\nabla\left[ \langle xy\rangle-\langle
x\rangle \langle y\rangle\right]=0$. Further, every independent
space must have conditional probabilities equal to marginal
probabilities and so satisfy $\nabla\left[P_{x|y}-P_x\right]=0$.
Finally, two independent variables have joint entropy equal to
the sum of the individual entropies so every independent space
must satisfy $\nabla\left[E_{xy}-E_{x}-E_{y}\right]=0$. These
relations evaluate differently in either ${\cal P}_{\rm ind}$
with isomorphism constraints or ${\cal P}'_{\rm ind}$ with limit
processes.  We have:
\begin{eqnarray}
 \nabla \left[P_{xy}(00)- P_x(0)P_y(0)\right]|_{ad=bc} &=& 0  \nonumber \\
 \lim_{ad\rightarrow bc} \nabla \left[P_{xy}(00)- P_x(0)P_y(0)\right]
     &=& \lim_{ad\rightarrow bc} \nabla (ad-bc) \neq 0 \nonumber \\
 \nabla \left[\langle xy\rangle-\langle x\rangle\langle y\rangle\right]|_{ad=bc}
   &=& 0  \nonumber \\
 \lim_{ad\rightarrow bc} \nabla \left[\langle xy\rangle-\langle x\rangle\langle y\rangle\right]
     &=& \lim_{ad\rightarrow bc}\nabla (ad-bc) \;\neq\; 0 \nonumber \\
 \nabla \left[P_{x|y}(0|0)-P_x(0)\right]|_{ad=bc} &=& 0  \nonumber  \\
 \lim_{ad\rightarrow bc} \nabla \left[P_{x|y}(0|0)-P_x(0)\right]
     &=& \lim_{ad\rightarrow bc}\nabla \left[\frac{ad-bc}{a+c} \right] \;\neq\; 0 \nonumber \\
 \nabla \left[E_{xy}-E_x-E_y\right]|_{ad=bc} &=& 0 \nonumber  \\
 \lim_{ad\rightarrow bc} \nabla \left[E_{xy}-E_x-E_y\right]
     &\neq &  0.
\end{eqnarray}
(See Eqs. \ref{eq_ind_space} to \ref{eq_ind_space2}.)

\begin{figure}[htb]
\centering
\includegraphics[width=0.9\columnwidth,clip]{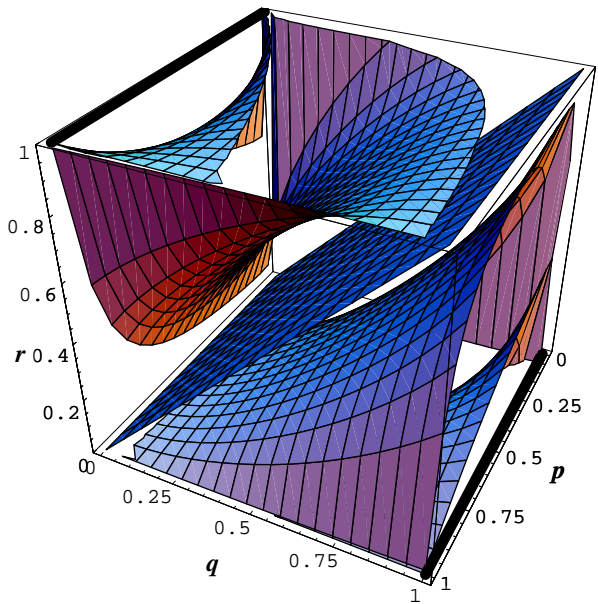}
\caption{\em  Every point within the $(p,q,r)$
probability space shown specifies a particular state of
correlation $\rho_{xy}(p,q,r)$ between the $x$ and $y$ variables.
We show here several lines and surfaces of constant correlation
taking values from top left to bottom right of
$\rho_{xy}=+1,+0.75,+0.25,0,-0.25,-0.75,-1$.  The optimization of
expectations at any point $(p,q,r)$ must take account of
correlated changes between $x$ and $y$.
 \label{f_correlation_constraints}}
\end{figure}

\subsection{Discussion}

There are two approaches to optimization over probability spaces
presented here.  Probability theory uses isomorphic constraints
to exactly preserve the properties of embedded probability
spaces and then compares these exactly calculated values. Game
theory eschews the use of isomorphic constraints and in effect,
argues that any uncertainty about which probability space to
choose bleeds into many calculations within a given space and
alters the calculated outcomes.

When probability spaces are represented as geometries, then it
is expected that at least some of the properties of the
probability space will be rendered in geometric terms.  How
these geometrical properties are preserved when a probability
space is embedded within another is the question.  Probability
theory requires the exact preservation of all properties of
every source space and this is achieved by imposing different
constraints on different points within the target space. Game
theory in contrast, imposes a single target space geometry onto
every source probability space. One way to picture this is shown
in Fig. \ref{f_tangent_spaces}. This figure shows how
probability theory exactly preserves the dimensionality and
tangent spaces of embedded probability spaces, while game theory
overwrites these properties of the embedded spaces with the
corresponding properties of the mixed space.

In probability theory, the different isomorphism constraints and
tangent spaces acting at each point define non-intersecting
lines and surfaces within the target space. Some of these are
shown in Fig. \ref{f_correlation_constraints} representing the
$(p,q,r)$ simplex of the two potentially correlated $x$ and $y$
variables (this behavioural space is defined in the next
section). Here, each state of correlation is a constant and
cannot vary during an optimization analysis so an optimization
procedure must sequentially take account of every possible
correlation state between these variables, setting
$\rho_{xy}=\rho$ for all $\rho\in[-1,1]$. These optimum points
can then be compared to determine which correlation state
between $x$ and $y$ returns the best value.

Unsurprisingly, these two distinct approaches can sometimes
generate conflicting results.

\begin{figure}[htb]
\centering
\includegraphics[width=\columnwidth,clip]{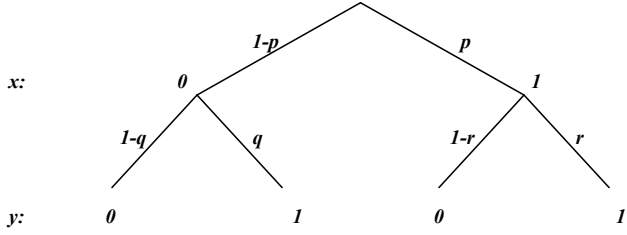}
\caption{\em  A simple decision tree where potentially independent
or correlated variables $x$ and $y$ take values $\{0,1\}$ with the
probabilities shown.   This defines the $(p,q,r)$ behavioural
probability space.
 \label{f_xy_decision_tree}}
\end{figure}

\section{Mixed and behavioural strategy spaces}
\label{sect_Mixed_and_behavioural_strategy_spaces}

The different approaches of probability theory and game theory
to isomorphic embeddings also impacts on the definitions of
mixed and behavioural strategy spaces. As usual, we will compare
these spaces both with and without isomorphism constraints. Our
focus will be on a simple decision problem involving two random
variables $x,y\in\{0,1\}$ where $y$ is potentially conditioned
on $x$ as shown in the behavioural strategy decision tree of
Fig. \ref{f_xy_decision_tree}.

\subsection{Mixed strategy space ${\cal P}_M$}

The mixed strategy space is denoted ${\cal P}_M$, and determines
the choice of $x$ via a probability distribution $\alpha$ while
the respective choices of $y$ on the left branch of the decision
tree $y_l$ and on the right branch $y_r$ are determined by an
independent probability distribution $\beta$ according to the
following table:
\begin{equation}                    \label{eq_mixed_strategies}
  \begin{array}{c|cccc}
 (y_l,y_r)=&   (0,0)    & (0,1)    &  (1,0)   &   (1,1)   \\\hline
     (x,y) &   \beta_0  & \beta_1  &  \beta_2 &   \beta_3 \\\hline
 \alpha_0  &    (0,0)   & (0,0)    &  (0,1)   &   (0,1)   \\
 \alpha_1  &    (1,0)   & (1,1)    &  (1,0)   &   (1,1).  \\
  \end{array}
\end{equation}
The mixed strategy simplex for each player is respectively
$S^X=\{(\alpha_0, \alpha_1)\in R_+^2: \sum_j \alpha_j=1\}$ and
$S^Y=\{(\beta_0, \beta_1, \beta_2, \beta_3)\in R_+^4: \sum_j
\beta_j=1\}$. The associated tangent spaces are $T^X=\{z\in R^2:
\sum_j z_j=0\}$ and $T^Y=\{z\in R^4: \sum_j z_j=0\}$, equivalent
to every possible positive or negative fluctuation in the
probabilities of the the pure strategies of each player. The
joint probability distribution $P_{xy}(x,y)$ for $x$ and $y$ is
\begin{eqnarray}               \label{eq_mixed_prob}
  P_{xy}(0,0) &=& (1-\alpha_1) (1 - \beta_2 - \beta_3) \nonumber \\
  P_{xy}(0,1) &=& (1-\alpha_1) (\beta_2 + \beta_3) \nonumber \\
  P_{xy}(1,0) &=& \alpha_1 (1 - \beta_1 - \beta_3) \nonumber \\
  P_{xy}(1,1) &=& \alpha_1 (\beta_1 + \beta_3).
\end{eqnarray}
Here, we have used normalization constraints to eliminate
$\alpha_0$ and $\beta_0$.  The expectations of the $x$ and $y$
variables are given by
\begin{eqnarray}                \label{eq_mixed_expec}
 \langle x\rangle &=& \alpha_1 \nonumber \\
 \langle y\rangle &=& \beta_2+\beta_3 +
                      \alpha_1(\beta_1-\beta_2) \nonumber \\
 \langle x y\rangle &=& \alpha_1 (\beta_1+\beta_3),
\end{eqnarray}
while their variances are
\begin{eqnarray}               \label{eq_mixed_var}
 V(x) &=& \alpha_1(1-\alpha_1) \nonumber \\
 V(y) &=& \left[\beta_2+\beta_3 + \alpha_1(\beta_1-\beta_2)\right]
           \times \nonumber \\
       && \times   \left[1-\beta_2-\beta_3 - \alpha_1(\beta_1-\beta_2)\right].
\end{eqnarray}
For completeness, we note the marginal and joint entropies are
\begin{eqnarray}
 E_{x}
   &=& - (1-\alpha_1)  \log(1-\alpha_1) - \alpha_1  \log\alpha_1  \nonumber \\
 E_{y}
   &=& - [1-\beta_2-\beta_3+\alpha_1(\beta_2-\beta_1)] \times \nonumber \\
   &&   \hspace{1cm} \log[1-\beta_2-\beta_3+\alpha_1(\beta_2-\beta_1)] \nonumber \\
   &&  -[\beta_2+\beta_3-\alpha_1(\beta_2-\beta_1)] \times \nonumber \\
   &&   \hspace{1cm}  \log[\beta_2+\beta_3-\alpha_1(\beta_2-\beta_1)] \nonumber \\
 E_{xy}
   &=& - (1-\alpha_1) (1-\beta_2-\beta_3) \log[(1-\alpha_1) (1-\beta_2-\beta_3)] \nonumber \\
   &&    - (1-\alpha_1) (\beta_2+\beta_3) \log[(1-\alpha_1) (\beta_2+\beta_3)] \nonumber \\
   &&  - \alpha_1 (1-\beta_1-\beta_3) \log[\alpha_1 (1-\beta_1-\beta_3)] \nonumber \\
   &&    - \alpha_1 (\beta_1+\beta_3) \log[\alpha_1 (\beta_1+\beta_3)].
\end{eqnarray}
Naturally, the mixed strategy probability space can model any
state of correlation between $x$ and $y$ with the correlation
give by
\begin{equation}                    \label{eq_mixed_correlation}
    \rho_{xy}(\alpha_1,\beta_1,\beta_2,\beta_3)
    = \frac{\sqrt{\alpha_1(1-\alpha_1)}(\beta_1-\beta_2)}{
    \sqrt{\langle y\rangle
    \left[1-\langle y\rangle\right]}}.
\end{equation}
Then, when $x$ and $y$ are perfectly correlated we have
$\rho_{xy}=1$ requiring the constraints $\beta_1=1$ and
$\beta_0=\beta_2=\beta_3=0$.  When $x$ and $y$ are perfectly
anti-correlated we have $\rho_{xy}=-1$ requiring the constraints
$\beta_2=1$ and $\beta_0=\beta_1=\beta_3=0$. Finally, when $x$
and $y$ are independent we have $\rho_{xy}=0$ requiring the
constraint $\beta_1=\beta_2$.

\begin{table*}
 \centering
\begin{ruledtabular}
\begin{tabular}{ccccc}
 $\rho_{xy}=1$ & ${\cal P}_M$  & ${\cal P}_B$  & $\left.{\cal P}_M\right|_{\beta_1=1}$ & $\left.{\cal P}_B\right|_{(q,r)=(0,1)}$ \\ \hline
 Parameters & $\alpha_1, \beta_1, \beta_2, \beta_3$ & $p, q, r$ & $\alpha_1$ & $p$ \\
 Dimensions & 4 & 3 & 1 & 1 \\
 $\nabla$ operator &
 $\frac{\partial}{\partial \alpha_1} \hat{\alpha}_1+\frac{\partial}{\partial \beta1} \hat{\beta}_1 + \frac{\partial}{\partial \beta_2} \hat{\beta}_2 + \frac{\partial}{\partial \beta_3} \hat{\beta}_3$ &
 $\frac{\partial}{\partial p} \hat{p}+\frac{\partial}{\partial q} \hat{q}+\frac{\partial}{\partial r} \hat{r}$ &
 $\frac{\partial}{\partial \alpha_1} \hat{\alpha}_1$ &
 $\frac{\partial}{\partial p} \hat{p}$ \\
  Gradient &
  $\lim_{\beta_1\rightarrow 1}\nabla (.)$ &
  $\lim_{(q,r)\rightarrow (0,1)}\nabla (.)$ &
  $\nabla$ &
  $\nabla$ \\
 \multicolumn{5}{l}{Probability Conservation}  \\
 $\nabla\left[P_{xy}(0,0)+P_{xy}(1,1)\right]$ &
 $ \alpha_1 \hat{\beta}_1 - (1-\alpha_1) \hat{\beta}_2+ (2\alpha_1 - 1) \hat{\beta}_3$ &
 $ - (1-p) \hat{q} + p \hat{r}$ &
 0 &
 0 \\
 $\nabla\left[P_{xy}(0,1)+P_{xy}(1,0)\right]$ &
 $ -\alpha_1 \hat{\beta}_1 + (1-\alpha_1) \hat{\beta}_2- (2\alpha_1 - 1) \hat{\beta}_3$ &
 $  (1-p) \hat{q} - p \hat{r}$ &
 0 &
 0 \\
 \multicolumn{5}{l}{Conditionals}  \\
 $\nabla P_{x|y}(0|0)$ &
 $\frac{\alpha_1}{1-\alpha_1} (\hat{\beta}_1 + \hat{\beta}_3)$ &
 $\frac{p}{1-p} \hat{r}$ &
 0 &
 0 \\
 $\nabla P_{x|y}(0|1)$ &
 $\frac{1-\alpha_1}{1\alpha_1} (\hat{\beta}_2+ \hat{\beta}_3)$ &
 $\frac{1-p}{p} \hat{q}$ &
 0 &
 0\\
 \multicolumn{5}{l}{Expectations}  \\
 $\nabla \langle x\rangle$ &
 $\hat{\alpha}_1$ &
 $\hat{p}$ &
 $\hat{\alpha}_1$ &
 $\hat{p}$ \\
 $\nabla \langle y\rangle$ &
 $\hat{\alpha}_1 + \alpha_1 \hat{\beta}_1 + (1-\alpha_1) \hat{\beta}_2 + \hat{\beta}_3$ &
 $\hat{p} + (1-p) \hat{q} + p \hat{r}$ &
 $\hat{\alpha}_1$ &
 $\hat{p}$\\
 $\nabla \langle x y\rangle$ &
 $\hat{\alpha}_1 + \alpha_1 \hat{\beta}_1 +\alpha_1 \hat{\beta}_3$ &
 $\hat{p} + p \hat{r}$ &
 $\hat{\alpha}_1$ &
 $\hat{p}$\\
 \multicolumn{5}{l}{Variance}  \\
 $\nabla \left[V(x)+V(y)-2\mbox{cov}(x,y)\right]$ &
 $-\alpha_1 \hat{\beta}_1+ (1-\alpha_1) \hat{\beta}_2+ (1-2\alpha_1)\hat{\beta}_3$ &
 $(1-p) \hat{q} - p \hat{r}$ &
 0 &
 0 \\
 \multicolumn{5}{l}{Entropy}  \\
  $\nabla\left[E_{xy}-E_{x}\right]$ & $\neq 0$ & $\neq 0$ &  0 & 0 \\
 \multicolumn{5}{l}{Correlation}  \\
  $\nabla \rho_{xy}$ & $\neq 0$ & $\neq 0$ &  0 & 0 \\
  &  &  &  &  \\
  &  &  &  & \\
  &  &  &  & \\   \hline \hline
 $\rho_{xy}=0$ & ${\cal P}_M$  & ${\cal P}_B$  & $\left.{\cal P}_M\right|_{\beta_1=\beta_2}$ & $\left.{\cal P}_B\right|_{r=q}$ \\  \hline
 Parameters & $\alpha_1, \beta_1, \beta_2, \beta_3$ & $p, q, r$ & $\alpha_1$, $\bar{\beta}=\beta_1+\beta_3$ & $p, q$ \\
 Dimensions & 4 & 3 & 2 & 2 \\
 $\nabla$ operator &
 $\frac{\partial}{\partial \alpha_1} \hat{\alpha}_1+\frac{\partial}{\partial \beta1} \hat{\beta}_1 + \frac{\partial}{\partial \beta_2} \hat{\beta}_2 + \frac{\partial}{\partial \beta_3} \hat{\beta}_3$ &
 $\frac{\partial}{\partial p} \hat{p}+\frac{\partial}{\partial q} \hat{q}+\frac{\partial}{\partial r} \hat{r}$ &
 $\frac{\partial}{\partial \alpha_1} \hat{\alpha}_1+\frac{\partial}{\partial \bar{\beta}} \hat{\bar{\beta}}$&
 $\frac{\partial}{\partial p} \hat{p}+\frac{\partial}{\partial q} \hat{q}$ \\
  Gradient &
  $\lim_{\beta_2\rightarrow \beta_1}\nabla  (.)$ &
  $\lim_{r\rightarrow q}\nabla (.)$ &
  $\nabla$ &
  $\nabla$ \\
 \multicolumn{5}{l}{Probability}  \\
 $\nabla \left[P_{xy}(0,0) - P_x(0)P_y(0)\right]$ &
 $\alpha_1 (1-\alpha_1) (\hat{\beta}_1 - \hat{\beta}_2)$ &
 $p (1-p) (\hat{r} - \hat{q})$ &
 0 &
 0\\
 $\nabla \left[P_{xy}(0,1) - P_x(0)P_y(1)\right]$ &
 $\alpha_1 (1-\alpha_1) (\hat{\beta}_2 - \hat{\beta}_1)$ &
 $p (1-p) (\hat{q} - \hat{r})$ &
 0 &
 0\\
 $\nabla \left[P_{xy}(1,0) - P_x(1)P_y(0)\right]$ &
 $\alpha_1 (1-\alpha_1) (\hat{\beta}_2 -  \hat{\beta}_1)$ &
 $p (1-p) (\hat{q} -  \hat{r})$ &
 0 &
 0\\
 $\nabla \left[P_{xy}(1,1) - P_x(1)P_y(1)\right]$ &
 $\alpha_1 (1-\alpha_1) (\hat{\beta}_1 - \hat{\beta}_2)$ &
 $p (1-p) (\hat{r} - \hat{q})$ &
 0 &
 0\\
 \multicolumn{5}{l}{Conditionals}  \\
 $\nabla \left[P_{x|y}(0|0)-P_x(0)\right]$ &
 $\frac{\alpha_1(1-\alpha_1)}{1-\beta_1-\beta_3} (\hat{\beta}_1-  \hat{\beta}_2)$ &
 $\frac{p(1-p)}{(1-q)} (\hat{r} - \hat{q})$ &
 0 &
 0\\
 $\nabla \left[P_{x|y}(0|1)-P_x(0)\right]$ &
 $\frac{\alpha_1(1-\alpha_1)}{\beta_1+\beta_3} (\hat{\beta}_2-\hat{\beta}_1)$ &
 $\frac{p(1-p)}{q} (\hat{q} -  \hat{r})$ &
 0 &
 0\\
 \multicolumn{5}{l}{Expectation}  \\
 $\nabla\left[\langle xy\rangle-\langle x\rangle \langle y\rangle\right]$ &
 $\alpha_1(1-\alpha_1) (\hat{\beta}_1 - \hat{\beta}_2)$ &
 $p(1-p) (\hat{r} -\hat{q})$ &
 0 &
 0\\
 \multicolumn{5}{l}{Entropy}  \\
  $\nabla\left[E_{xy}-E_{x}-E_{y}\right]$ & $\neq 0$ & $\neq 0$ &  0 & 0 \\
 \multicolumn{5}{l}{Correlation}  \\
 $\nabla \rho_{xy}$ & $\neq 0$ & $\neq 0$ &
 0 &
 0\\
  &  &  &  & \\
  &  &  &  & \\
\end{tabular}
\end{ruledtabular}
 \caption{\em A comparison of calculated results for mixed ${\cal P}_M$
 and behavioural ${\cal P}_B$ strategy spaces with those same spaces when
 subject to isomorphic constraints.  We examine points where respectively
 the $x$ and $y$ variables are first perfectly correlated with $\rho_{xy}=1$
 and then independent with $\rho_{xy}=0$. In the unconstrained behavioural spaces,
 all quantities are evaluated at points satisfying $\lim_{\beta_1 \rightarrow 1}$ or
 $\lim_{(q,r) \rightarrow (0,1)}$ when $\rho_{xy}=1$, and at points satisfying
 $\lim_{\beta_2 \rightarrow \beta_1}$ or  $\lim_{r\rightarrow q}$ when $\rho_{xy}=0$.
 The isomorphically constrained spaces are respectively
 indicated by $\left.{\cal P}_M\right|_{\beta_1=1}$ and $\left.{\cal P}_B\right|_{(q,r)=(0,1)}$
 for the perfectly correlated case, and
 $\left.{\cal P}_M\right|_{\beta_1=\beta_2}$ and $\left.{\cal
 P}_B\right|_{r=q}$ when the variables are independent.
 Game theory and probability theory assign
 different dimensionality and tangent spaces to these cases.   Many
 calculated results differ between these spaces. }
 \label{t_tangent_space_effects}
\end{table*}

\subsection{Behavioural strategy space  ${\cal P}_B$}

The behavioural strategy probability space \cite{Kuhn_1953} is
denoted ${\cal P}_B$ and is parameterized as shown in Fig.
\ref{f_xy_decision_tree}. The behavioural strategy space for the
players is $S^{XY}=\{(p,q,r)\in R_+^3: 0\leq p, q, r \leq 1\}$
after taking account of normalization. The associated tangent
space is $T^{XY}=\{z\in R^3\}$. The probability $P_{xy}(x,y)$
that $x$ and $y$ take on their respective values is
\begin{eqnarray}             \label{eq_behav_prob}
  P_{xy}(0,0) &=& (1-p)(1-q) \nonumber \\
  P_{xy}(0,1) &=& (1-p)q \nonumber \\
  P_{xy}(1,0) &=& p(1-r) \nonumber \\
  P_{xy}(1,1) &=& pr.
\end{eqnarray}
This distribution gives the following expected values:
\begin{eqnarray}             \label{eq_behav_expect}
 \langle x\rangle &=& p \nonumber \\
 \langle y\rangle &=& q + p (r - q) \nonumber \\
 \langle x y\rangle &=& p r,
\end{eqnarray}
while the variances of the $x$ and $y$ variables are
\begin{eqnarray}              \label{eq_behav_var}
 V(x) &=& p(1-p) \nonumber \\
 V(y) &=& \left[q + p (r - q)\right]\left[1-q - p (r - q)\right].
\end{eqnarray}
The marginal and joint entropies between the $x$ and $y$
variables are
\begin{eqnarray}
 E_{x}
   &=& - (1-p) \log(1-p) - p \log p \nonumber \\
 E_{y}
   &=& - [(1-p) (1-q)+p (1-r)] \times \nonumber \\
   && \hspace{1cm} \log[(1-p) (1-q)+p (1-r)] \nonumber \\
   &&    - [(1-p) q+p r] \log[(1-p) q+p r] \nonumber \\
 E_{xy}
   &=& - (1-p) (1-q) \log[(1-p) (1-q)] \nonumber \\
   &&    - (1-p) q \log[(1-p) q] \nonumber \\
   &&  - p (1-r) \log[p (1-r)] \nonumber \\
   &&    - p r \log[pr].
\end{eqnarray}
The behavioural probability space also allows modeling any
arbitrary state of correlation between the $x$ and $y$ variables
where the correlation between $x$ and $y$ is
\begin{equation}           \label{eq_rho_correlation}
    \rho_{xy}
    = \frac{\sqrt{p(1-p)}(r-q)}{
        \sqrt{\left[q+p(r-q)\right]\left[1-q-p(r-q)\right]}}.
\end{equation}
Then, $x$ and $y$ are perfectly correlated at
$\rho_{xy}(p,0,1)=1$, perfectly anti-correlated at
$\rho_{xy}(p,1,0)=-1$, and uncorrelated if either $p=0$ or $p=1$
or $q=r$ giving $\rho_{xy}=0$. Hence, the decision tree of Fig.
\ref{f_xy_decision_tree} encompasses every possible state of
correlation between $x$ and $y$, and thus it can be used to
perform a complete analysis.

\subsection{Isomorphic Mixed and Behavioural Spaces}

The mixed ${\cal P}_M$ and behavioural ${\cal P}_B$ strategy
spaces contain embedded probability spaces where $x$ and $y$ are
respectively perfectly correlated, independent, or partially
correlated. As previously, we will now perform a comparison of
probability spaces, both with and without isomorphic
constraints, for various correlation states between the $x$ and
$y$ variables.  That is, we will compare the mixed strategy
space ${\cal P}_M$ and behavioural strategy space ${\cal P}_B$
with isomorphism constrained mixed and behavioural strategy
spaces as indicated using the following notation.

The case of perfectly correlated $x$ and $y$ variables is
modeled by the spaces
\begin{equation}
\begin{array}{ll}
  \lim_{\beta_1\rightarrow 1} {\cal P}_M   &     {\rm mixed}   \\
  \left.{\cal P}_M\right|_{\beta_1=1}      &     {\rm constrained \; mixed}    \\
  \lim_{(q,r)\rightarrow (0,1)} {\cal P}_B &     {\rm behavioural}   \\
  \left.{\cal P}_B\right|_{(q,r)=(0,1)}    &     {\rm constrained \; behavioural}    \\
\end{array}
\end{equation}
In these spaces we expect all of the following to hold:
\begin{itemize}
\item $\nabla\left[P_{xy}(0,0)+P_{xy}(1,1)\right]=0$,
\item $\nabla\left[P_{xy}(0,1)+P_{xy}(1,0)\right]=0$,
\item $\nabla\left[P_{x|y}(0|0)\right]=0$,
\item $\nabla\left[P_{x|y}(0|1)\right]=0$,
\item $\nabla\left[\langle x\rangle - \langle
    y\rangle\right]=0$
\item $\nabla\left[\langle x\rangle - \langle
    xy\rangle\right]=0$
\item $\nabla\left[\langle y\rangle - \langle
    xy\rangle\right]=0$
\item
    $\nabla[V(x-y)]=\nabla\left[V(x)+V(y)-2\mbox{cov}(x,y)\right]=0$
\item $\nabla\left[E_{xy}-E_{x}\right]=0$.
\end{itemize}
Alternately, when $x$ and $y$ are independent, the relevant
spaces are
\begin{equation}
\begin{array}{ll}
  \lim_{\beta_1\rightarrow \beta_2} {\cal P}_M   &     {\rm mixed}   \\
  \left.{\cal P}_M\right|_{\beta_1=\beta_2}      &     {\rm constrained \; mixed}    \\
  \lim_{r\rightarrow q} {\cal P}_B               &     {\rm behavioural}   \\
  \left.{\cal P}_B\right|_{r=q}                  &     {\rm constrained \; behavioural}    \\
\end{array}
\end{equation}
In all these spaces, the probability distributions satisfy
\begin{itemize}
\item $\nabla\left[P_{xy}-P_xP_y\right]=0$
\item $\nabla\left[P_{x|y}-P_x\right]=0$
\item $\nabla\left[\langle xy\rangle-\langle x\rangle
    \langle y\rangle\right]=0$
\item $\nabla\left[E_{xy}-E_{x}-E_{y}\right]=0$.
\end{itemize}

Table \ref{t_tangent_space_effects} records whether each of the
expected relations is satisfied for each of the mixed and
behavioural spaces when they are either unconstrained, or
isomorphically constrained. As might be expected, the results
indicate that the weak isomorphisms used to construct the mixed
and behavioural spaces of game theory are not able to reproduce
necessarily true results from probability theory. Hence, the
rational player of game theory is unable to reliably reproduce
results from probability theory. These differences between game
theory and probability theory need to be resolved.

\section{Optimizing simple decision trees}
\label{sect_Optimizing_simple_decision_trees}

We now turn to consider how the differences between probability
theory and game theory influence decision tree optimization. We
consider the usual two potentially correlated random variables
depicted in Fig. \ref{f_xy_decision_tree} and will use both the
unconstrained behavioural probability space ${\cal P}_B$ and the
isomorphically constrained behavioural spaces $\left.{\cal
P}_B\right|_{\rho_{xy}=\rho}$ for every value of the correlation
state $\rho\in[-1,1]$.  Our goal is to present an optimization
problem in which a rational player following the rules of game
theory cannot achieve the payoff outcomes of a player following
the rules of probability theory. We suppose that a player gains
a payoff by advising a referee of the parameters of the decision
tree probability space $(p,q,r)$ to optimize a given nonlinear
random function. The referee uses these parameters to determine
the value of the function and provides a payoff equivalent to
this value.  (If desired, the referee could estimate the
probability parameters by using indicator functions and
observing an ensemble average of decision tree outcomes.)

There are many possible random functions which we could use, and
some are listed in Table \ref{t_tangent_space_effects}.  We
could choose any relation of the form $f=0$ where probability
theory shows $\nabla f=0$ and game theory has $\nabla f\neq 0$.
Therefore $\nabla f$ is effectively a discrepancy vector. We
focus on the squared magnitude of the length of the discrepancy
vector and examine functions of the form $F=1-|\nabla f|^2$.
Immediately, probability theory will optimize this function at
the point $F=1$ while game theory will locate an optimum at
$F<1$.  In particular, we choose
\begin{equation}
   f=P_{xy}(0,0)+P_{xy}(1,1)
\end{equation}
so
\begin{eqnarray}
   F &=& 1-\big|\nabla \left[P_{xy}(0,0)+P_{xy}(1,1)\right]\big|^2 \nonumber \\
     &=& 1-\big|\nabla \left[1-q+p(q+r-1)\right]\big|^2.
\end{eqnarray}

In the unconstrained behavioural space ${\cal P}_B$, a rational
player will evaluate this as
\begin{equation}
   F = 1-(1-q-r)^2-(1-p)^2-p^2.
\end{equation}
In turn, this will be maximized at points $p=\frac{1}{2}$ and
$q+r=1$ to give a maximum payoff of $F_{\rm max}=\frac{1}{2}$.

A contrasting result is obtained using the isomorphism
constraints of probability theory where our player faces the
optimization problem
\begin{eqnarray}
 \max F &=& 1-\big|\nabla \left[1-q+p(q+r-1)\right]\big|^2 \nonumber \\
     && \hspace{-1cm}
     \mbox{ subject to } \rho_{xy}=\rho, \;\;
                         \forall \rho\in[-1,1].
\end{eqnarray}
Our player might commence by adopting the constraint
$\rho_{xy}=1$ implemented by $(q,r)=(0,1)$ to give
\begin{eqnarray}
   \max F
      &=& \left. 1-\big|\nabla \left[1-q+p(q+r-1)\right]\big|^2 \right|_{(q,r)=(0,1)} \nonumber \\
      &=& 1.
\end{eqnarray}
This analysis leads to an optimum point at arbitrary $p$ and
$(q,r)=(0,1)$ and a maximum payoff of $F_{\rm max}=1$.
Self-evidently, the player would cease their optimization
analysis at this point as the achieved maximum can't be
improved.

Of course, there are many random functions defined over decision
trees which produce identical results when using or not using
isomorphic constraints.  We now briefly illustrate this using
polylinear expected payoff functions, and consider optimizing
the function
\begin{eqnarray}
 \max \langle\Pi\rangle &=& 2 \langle x\rangle + 3 \langle y\rangle
                       - 4 \langle x\rangle \langle y\rangle. \nonumber \\
     && \hspace{-1cm}
     \mbox{ subject to } \rho_{xy}=\rho, \;\;
                         \forall \rho\in[-1,1]
\end{eqnarray}
over the decision tree of Fig. \ref{f_xy_decision_tree}. Of
course, simple inspection will locate the optimum at $(\langle
x\rangle,\langle y\rangle)=(0,1)$ giving an expected payoff of
$\langle\Pi\rangle=3$.  However, we step through the process for
later generalization to strategic games.

There are an infinite number of correlation constraints to be
examined, but several are straightforward.   When the variables
are perfectly correlated at $\rho_{xy}=1$ via the constraint
$(q,r)=(0,1)$, we have $\langle x\rangle=\langle
y\rangle=\langle xy\rangle$ giving
\begin{equation}
 \langle\Pi\rangle = \langle x\rangle.
\end{equation}
This is optimized by setting $\langle x\rangle=1$ giving an
expected payoff of $\langle\Pi\rangle=1$.  Conversely, when
$\rho_{xy}=0$ and $x$ and $y$ are independent as occurs when
using the constraint $r=q$, then the expectations are separable
giving $\langle xy\rangle=\langle x\rangle\langle y\rangle$ and
\begin{equation}
 \langle\Pi\rangle = 2 \langle x\rangle + 3 \langle y\rangle
                       - 4 \langle x\rangle \langle y\rangle.
\end{equation}
As the $\langle x\rangle$ and $\langle y\rangle$ variables are
independent, a check of internal stationary points and the
boundary leads to an optimal point at $(\langle x\rangle,\langle
y\rangle)=(0,1)$ and an expected payoff of
$\langle\Pi\rangle=3$.

More general correlation states require use of, for instance,
standard Lagrangian optimization procedures. However, we here
adopt a numerical optimization approach by first using the
correlation constraint to write the $r$ variable as a function
of $p$, $q$ and the correlation constant $\rho$,
$r=r_{+}(p,q,\rho)$---see Eq. \ref{eq_r_plus_minus}. The
constraint $0\leq r\leq 1$ places limits on the permissible
values of $(p,q)$ and these are detailed in Eqs.
\ref{eq_permissible_region1} and \ref{eq_permissible_region2}.
The problem is then solved using a a typical Mathematica command
line of \cite{Pinter_2012}
{\bf%
\begin{eqnarray}
   && \mbox{NMaximize}[\{\mbox{inRange}[r_+(p,q,\rho)] \times \nonumber \\
  && \left[2p+3q-3pq-p r_+(p,q,\rho)\right], \nonumber \\
   && 0\leq p\leq 1 \mbox{ \&\& } 0\leq q\leq 1
       \},\{p,q\}].
\end{eqnarray}
}%
Here, a suitably defined ``inRange" function determines whether
$r_+$ is taking permissible values between zero and unity
allowing the payoff function to be examined over the entire
$(p,q)$ plane. The resulting optimal expected payoffs follow:
\begin{equation}
  \begin{array}{l|l|l|l}
  \rho&          (p,q,r)               & \langle\Pi\rangle \\ \hline
    +1      &  (1.,0.,1.)              &  1.               \\
    +0.75   &  (0.8138,0.3876,1.)      &  1.03032          \\
    +0.5    &  (0.4831,0.5917,1.)      &  1.40068          \\
    +0.25   &  (0.2590,0.7953,1.)      &  2.02693          \\
     0      &  (0.,1.,1.)              &  3.               \\
    -0.25   &  (0.,1.,0.9378)          &  3.               \\
    -0.5    &  (0.,1.,0.7506)          &  3.               \\
    -0.75   &  (0.,1.,0.4386)          &  3.               \\
    -1      &  (0.,1.,0.)              &  3.           \\
  \end{array}
\end{equation}
Some care must be taken to ensure convergence of the solutions.
This analysis makes it evident that the player can maximize
expected payoffs by choosing a correlation constraint where $x$
and $y$ is independent (say) allowing the setting
$(p,q,r)=(0,1,1)$ to gain a payoff of $\langle\Pi\rangle=3$.
Other choices would also have been possible.

We now turn to applying isomorphism constraints to the strategic
analysis of game theory.

\section{Optimizing a multistage game tree}
\label{sect_Optimizing_a_multistage_game_tree}

In this section, we show that the use of isomorphic constraints
can alter the outcomes of strategic games even when expected
payoff functions are being used.  As usual, we will consider
either the behavioural strategy space ${\cal P}_B$ (Eq.
\ref{eq_behav_prob}) or the isomorphically constrained
behavioural spaces $\left.{\cal P}_B\right|_{\rho_{xy}=\rho}$
for every value of the correlation state $\rho\in[-1,1]$.

We consider a strategic interaction between two players over
multiple stages as depicted in the behavioural strategy space of
Fig. \ref{f_xy_decision_tree}. Here, two players denoted $X$ and
$Y$ seek to optimize their respective payoffs
\begin{eqnarray}
 X: \max \Pi^{X}(x,y) &=& 3 - 2 x - y + 4 x y \nonumber \\
 Y: \max \Pi^{Y}(x,y) &=& 1 + 3 x + y - 2 x y.
\end{eqnarray}
We assume that player $X$ chooses the value of $x$ and advises
this to $Y$ before $Y$ determines the value of $y$.

In the unconstrained behavioural strategy space ${\cal P}_B$,
this perfect information game is optimized using backwards
induction to give the pure strategy choices $(x,y)=(0,1)$
achieving payoffs of $(\Pi^X,\Pi^Y)=(2,2)$.

We now consider the constrained behavioural spaces $\left.{\cal
P}_B\right|_{\rho_{xy}=\rho}, \forall \rho\in[-1,1]$. The two
players are non-communicating and it is generally not possible
to use a single value for the correlation $\rho$, and this
generally makes the analysis intractable. However, player $Y$
has total control over the setting of the correlation $\rho$ in
three cases---when $\rho=\pm 1$ and $\rho=0$.  We consider these
cases now. First consider the space ${\cal
P}_{B}|_{\rho_{xy}=1}$ in which the variables are functionally
equal so $y=x=xy$. In this space the players face the respective
optimization tasks
\begin{eqnarray}
 X: \max_x \Pi^{X}(x) &=& 3 +  x \nonumber \\
 Y:        \Pi^{Y}(x) &=& 1 +  2x.
\end{eqnarray}
As a result, player $X$ optimizes their payoff by setting $x=1$
giving the outcomes $(\Pi^{X},\Pi^{Y})=(4,3)$. In contrast, in
the space ${\cal P}_{B}|_{\rho_{xy}=-1}$, the variables are
functionally related by $y=1-x$ and $xy=0$.  These constraints
render the optimization tasks as
\begin{eqnarray}
 X: \max_x \Pi^{X}(x) &=& 2 -  x \nonumber \\
 Y:        \Pi^{Y}(x) &=& 2 + 2 x.
\end{eqnarray}
Here, player $X$ chooses $x=0$ to optimize their payoff leading
to the outcomes $(\Pi^{X},\Pi^{Y})=(2,2)$.  Finally, when player
$Y$ chooses to discard all information about the $x$ variable,
then the variables $x$ and $y$ are independent and the chosen
space is ${\cal P}_{B}|_{\rho_{xy}=0}$.  In this space, there
are no pure strategy solutions and the players will optimize
expected payoffs.  We have $\langle x\rangle=p$ and $\langle
y\rangle=q$ and $\langle xy\rangle=\langle x\rangle\langle
y\rangle=pq$ giving the optimization problem
\begin{eqnarray}
 X:\max_p \langle\Pi^X\rangle &=&
     3 - 2 p - q + 4 p q \nonumber \\
 Y:\max_q \langle\Pi^Y\rangle &=&
     1 + 3 p + q - 2 p q.
\end{eqnarray}
The best response functions or equivalent partial differentials
are
\begin{eqnarray}
 X:  \frac{\partial \langle\Pi^X\rangle}{\partial p}  &=&
     - 2 + 4  q \nonumber \\
 Y: \frac{\partial \langle\Pi^Y\rangle}{\partial q}  &=&
     1 - 2 p
\end{eqnarray}
locating the optimal point at $(p,q)=(\frac{1}{2},\frac{1}{2})$
with expected payoffs of
$(\langle\Pi^X\rangle,\langle\Pi^Y\rangle)=(\frac{5}{2},\frac{5}{2})$.

At this stage of the analysis, both players have separately
calculated an equilibrium point in three spaces ${\cal
P}_{B}|_{\rho_{xy}=\rho}$ for $\rho\in\{-1,0,1\}$, and the
selection of these correlation states is solely at the
discretion of player $Y$. The expected payoffs gained at each of
these ``local" equilibrium points can then be compared to obtain
a ``global" optimal expected payoff. For convenience, these are
summarized here:
\begin{equation}
 \begin{array}{cc}
   \rho &  (\langle\Pi^X\rangle,\langle\Pi^Y\rangle) \\ \hline
       -1       &     (2,2)                          \\
       0        &     (\frac{5}{2},\frac{5}{2})      \\
       +1       &     (4,3).                         \\
 \end{array}
\end{equation}
Based on these results, player $Y$ will then rationally optimize
their expected payoff by choosing to have their variables in a
state of perfect correlation with $\rho=1$ in the space ${\cal
P}_{B}|_{\rho_{xy}=1}$.  Player $X$, also being a rational
optimizer will play accordingly to give equilibrium payoffs of
$(\langle\Pi^X\rangle,\langle\Pi^Y\rangle)=(4,3)$.

As noted above, the more general treatment of a strategic game,
even one as simple as this one, appears intractable.

\section{Conclusion}

A rational player must compare expected payoffs across the mixed
strategy space in order to locate equilibria.  As expectations
are polylinear, such comparisons are mathematically equivalent
to calculating gradients and the issues raised in this paper
apply. Further, it is perfectly possible that rational player
might need to calculate the Fisher information defined in terms
of gradients of probability distributions in order to optimize
payoffs. It is perfectly possible that a rational player might
need to optimize an Entropy gradient to maximize a payoff. It is
even possible to define games where payoffs depend directly on
the gradient of a probability distribution---shine light through
glass sheets painted by players to alter transmission
probabilities and make payoffs dependent on the resulting light
intensity gradients (call it the interior decorating game). This
paper has shown that rational players working with the standard
strategy spaces of game theory will have difficulties with these
games.

This paper has highlighted two alternate ways to optimize a
multivariate function $\Pi(x,y)$ where $x$ and $y$ might be
functionally related in different ways, $y=g_i(x)$ for different
$i$ say. The first approach, common to probability theory and
general optimization theory, considers each potential functional
relation as occupying a distinct space and approaches the
optimization as a choice between distinct spaces. Any
uncertainty about which space to choose does not leak into the
properties of any individual space.  If desired, isomorphic
constraints can be used to embed all these distinct spaces into
a single enlarged space for convenience, but if so, all the
properties of the optimization problem are exactly preserved.
The second approach, common to game theory, holds that the
uncertainty about which functional relation to choose should
appear in the same space as the variables $(x,y)$. This is
accomplished by expanding the size of the space to include both
the old variables $x$ and $y$ and sufficient new variables (not
explicitly shown here) to contain all the potential functional
relations and allow $\lim_{y\rightarrow
g_i(x)}\Pi(x,y)=\Pi[x,g_i(x)]$ for all $i$. This enlarged space
then allows gradient comparisons to be made at points
$\Pi[x,g_i(x)]-\Pi[x,g_j(x)]$ for all $i$ and $j$ to locate
optima. These two approaches can lead to conflicting
optimization outcomes as while these approaches generally assign
the same values to functions at all points,
\begin{equation}
  \left.\Pi(x,y)\right|_{y=g_i(x)} =
     \lim_{y\rightarrow g_i(x)} \Pi(x,y),
\end{equation}
they typically calculate different gradients at those same
points
\begin{equation}
  \left.\nabla \Pi(x,y)\right|_{y=g_i(x)} \neq
     \lim_{y\rightarrow g_i(x)} \nabla \Pi(x,y).
\end{equation}
These differences can be extreme when the function $\Pi(x,y)$
depends on global properties of the space---the dimension,
volume, gradient, information or entropy say.  In its approach,
game theory differs from many other fields including other
fields of economics.  For example, the Euler-Lagrange equations
of Ramsey-type models consider the functional variation of some
function $u[y(x),y'(x)]$ while ensuring a consistent treatment
of the function $y(x)$ and its gradient $y'(x)$
\cite{Ramsey_1928_543}. Gradients are not taken in limits in
these fields.

Throughout this paper, we have presumed that a rational player
should be able to use standard techniques from either
probability theory or optimization theory on the one hand, or
decision theory and game theory on the other, and expect all of
these methods to provide consistent results. We have shown that
when considering multiple, potentially correlated variables, and
functions of these variables dependent on the geometry of the
probability parameter space, then these methods can give rise to
contradictory optimization outcomes.  We have suggested decision
and game theory are incomplete when they require the adoption of
a single geometry for any decision or game tree, and that these
fields should consider applying the alternate geometries of
probability theory and optimization theory.  Recognizing that a
single multi-stage decision or game tree can encompass an
infinite number of incommensurate probability spaces might
resolve some of the paradoxes of game theory, and have broader
application.

\section{Acknowledgments}
The author gratefully acknowledges discussions with Kae Nemoto.

\bibliographystyle{unsrt}

\appendix

\section{Optimization and isomorphic probability spaces}
\label{app_Optimization_and_isomorphic_probability_spaces}

\subsection{Isomorphic dice}

In each respective die space, the gradient operator is
\begin{equation}
   \nabla =   \sum_{i=1}^{n-1}  \hat{p}_i \frac{\partial}{\partial p_i}
\end{equation}
where a hatted variable $\hat{p}_i$ is a unit vector in the
indicated direction and we resolve the normalization constraint
via $p_n=1-\sum_{i=1}^{n-1}p_i$. For the coin, we have
\begin{eqnarray}               \label{eq_coin_functions}
  V &=& \int_{0}^{1} da \int_{0}^{1} db \; \delta_{a+b=1} \nonumber \\
    &=& \int_{0}^{1}  da \nonumber \\
    &=& 1      \nonumber \\
  E_x &=&  -[ a \log(a) + (1-a) \log(1-a)] \nonumber \\
  \nabla E_x &=& - \hat{a} \log \frac{a}{1-a}.
\end{eqnarray}
For the triangle, the equivalent functions are
\begin{eqnarray}                \label{eq_triangle_functions}
  V &=& \int_{0}^{1} da \int_{0}^{1} db \int_{0}^{1} dc \; \delta_{a+b+c=1} \nonumber \\
    &=& \int_{0}^{1}  da \int_{0}^{1-a}  db \nonumber \\
    &=& \frac{1}{2}      \nonumber \\
  E_x &=&  -[ a \log(a) + b \log(b) + \nonumber  \\
      &&    \hspace{1cm} (1-a-b) \log(1-a-b)] \nonumber \\
  \nabla E_x &=& - \hat{a} \log \frac{a}{1-a-b} - \hat{b} \log \frac{b}{1-a-b}.
\end{eqnarray}
Finally, for the square, we have
\begin{eqnarray}               \label{eq_square_functions}
  V &=& \int_{0}^{1} da \int_{0}^{1} db \int_{0}^{1} dc  \int_{0}^{1} dd \; \delta_{a+b+c+d=1} \nonumber \\
    &=& \int_{0}^{1}  da \int_{0}^{1}  db \int_{0}^{1-a-b}  dc \nonumber \\
    &=& \frac{1}{6}      \nonumber \\
  E_x &=&  -[ a \log(a) + b \log(b) + c \log(c) + \nonumber \\
       &&  \hspace{1cm} (1-a-b-c) \log(1-a-b-c)] \nonumber \\
  \nabla E_x &=& - \hat{a} \log \frac{a}{1-a-b-c} -
                   \hat{b} \log \frac{b}{1-a-b-c}   \nonumber \\
      &&  \hspace{1cm} - \hat{c} \log \frac{c}{1-a-b-c}.
\end{eqnarray}

The function $F(a,b,c)$ has a directed gradient in the direction
$\frac{1}{\sqrt{2}}(1,-1,0)$ of
\begin{equation}
  \nabla F(a,b,c) . \frac{1}{\sqrt{2}}(1,-1,0)
   = V^2 \frac{1}{2} \log \frac{b}{a}
\end{equation}
using Eq. \ref{eq_square_functions}.  At points where
$(a,b,c)=(a,1-a,0)$ this gives a directed gradient of
\begin{equation}       \label{eq_directed_grad}
  \nabla F(a,1-a,0) . \frac{1}{\sqrt{2}}(1,-1,0)
   = V^2 \frac{1}{2} \log \frac{1-a}{a}
\end{equation}
which is optimized at $(a,b,c)=(\frac{1}{2},\frac{1}{2},0)$.

\subsection{Continuous bivariate Normal spaces}

Two continuous independent and normally distributed random
variables $x$ and $y$ with respective means $\mu_x$ and $\mu_y$
and standard deviations $\sigma_x$ and $\sigma_y$ have joint and
marginal distributions of
\begin{eqnarray}     \label{eq_bivariate_normal_ind}
  P_{xy} &=& \frac{1}{2\pi\sigma_x\sigma_y}
    e^{-\frac{1}{2}
    \left[ \frac{(x-\mu_x)^2}{\sigma_x^2} +
      \frac{(y-\mu_y)^2}{\sigma_y^2} \right] } \nonumber \\
  P_{x} &=& \frac{1}{\sqrt{2\pi}\sigma_x}
    e^{-\frac{1}{2}
    \frac{(x-\mu_x)^2}{\sigma_x^2} } \nonumber \\
  P_{y} &=& \frac{1}{\sqrt{2\pi}\sigma_y}
    e^{-\frac{1}{2}
    \frac{(y-\mu_y)^2}{\sigma_y^2} }.
\end{eqnarray}
The conditional distribution for $x$ given some value of $y$ is
\begin{equation}
  P_{x|y} = \frac{1}{\sqrt{2\pi}\sigma_x}
    e^{-\frac{1}{2}
     \frac{(x-\mu_x)^2}{\sigma_x^2} }.
\end{equation}
Two random normally distributed variables $x$ and $y$ with
correlation value $\rho$ have a joint distribution
\begin{eqnarray}            \label{eq_bivariate_normal_corr}
  P'_{xy} &=& \frac{1}{2\pi\sigma_x\sigma_y \sqrt{1-\rho^2}} \times  \\
    && e^{-\frac{1}{2(1-\rho^2)}
    \left[ \frac{(x-\mu_x)^2}{\sigma_x^2}
     - \frac{2 \rho (x-\mu_x) (y-\mu_y)}{\sigma_x \sigma_y} +
      \frac{(y-\mu_y)^2}{\sigma_y^2} \right] }. \nonumber
\end{eqnarray}
The marginal distributions for the correlated case are identical
to those of the independent space so $P'_{x}=P_{x}$ and
$P'_{y}=P_{y}$. The conditional distribution for $x$ given some
value of $y$ is
\begin{equation}
  P'_{x|y} = \frac{1}{\sqrt{2\pi(1-\rho^2)}\sigma_x}
    e^{-\frac{1}{2(1-\rho^2)}
     \frac{(x-\bar{\mu}_x)^2}{\sigma_x^2} },
\end{equation}
where the new conditioned mean is
\begin{equation}
  \bar{\mu}_x=\mu_x+\rho \frac{\sigma_x}{\sigma_y} (y - \mu_y).
\end{equation}
In the enlarged distribution space, the gradient operator is
\begin{eqnarray}
 \nabla &=& \frac{\partial}{\partial x} \hat{x}
           + \frac{\partial}{\partial y} \hat{y}
           + \frac{\partial}{\partial \mu_x} \hat{\mu}_x
           + \frac{\partial}{\partial \mu_y} \hat{\mu}_y +
           \nonumber \\
    &&     \frac{\partial}{\partial \sigma_x} \hat{\sigma}_x
           + \frac{\partial}{\partial \sigma_y} \hat{\sigma}_y
           + \frac{\partial}{\partial \rho} \hat{\rho}.
\end{eqnarray}
When suitably constrained by an isomorphism, the enlarged
distribution satisfies
\begin{eqnarray}
   \left. \nabla \left[P'_{xy}-P'_xP'_y\right]\right|_{\rho=0}
         &=& 0 \nonumber \\
   \left. \nabla \left[P'_{x|y}-P'_x\right]\right|_{\rho=0}  &=& 0.
\end{eqnarray}
Conversely, when the parameter $\rho$ is not externally
constrained then these required relations are not held even in
the limit as $\rho\rightarrow 0$ as
\begin{eqnarray}        \label{eq_bivariate_normal_corr_t2}
   \lim_{\rho\rightarrow 0} \nabla \left[P'_{xy}-P'_xP'_y\right]
     &=& \hat{\rho} \lim_{\rho\rightarrow 0} \frac{\partial}{\partial \rho} P'_{xy} \neq 0  \nonumber \\
   \lim_{\rho\rightarrow 0} \nabla \left[P'_{x|y}-P'_x\right]
      &=& \hat{\rho} \lim_{\rho\rightarrow 0} \frac{\partial}{\partial \rho} P'_{x|y} \neq 0.
\end{eqnarray}
Expectations of the $x$ and $y$ variables must also satisfy
certain gradient relations.  As expectations integrate over the
$x$ and $y$ variables, the gradient operator is a function of
only five variables now,
\begin{equation}
 \nabla = \frac{\partial}{\partial \mu_x} \hat{\mu}_x
           + \frac{\partial}{\partial \mu_y} \hat{\mu}_y +
            \frac{\partial}{\partial \sigma_x} \hat{\sigma}_x
           + \frac{\partial}{\partial \sigma_y} \hat{\sigma}_y
           + \frac{\partial}{\partial \rho} \hat{\rho}.
\end{equation}
We then have
\begin{eqnarray} \label{eq_bivariate_normal_corr_t3}
   \left. \nabla \left[ \langle xy\rangle'-\langle x\rangle' \langle y\rangle'\right]\right|_{\rho=0}
        &=& 0  \\
   \lim_{\rho\rightarrow 0}  \nabla \left[ \langle xy\rangle'-\langle x\rangle' \langle y\rangle'\right]
       &=& \hat{\rho} \lim_{\rho\rightarrow 0} \frac{\partial}{\partial \rho} \langle xy\rangle' \neq 0.\nonumber
\end{eqnarray}

\subsection{Joint probability space optimization}

The gradient operator in the probability space of the square
dice with probability parameters $(a,b,c)$ is
\begin{equation}
 \nabla = \hat{a} \frac{\partial }{\partial a} +
             \hat{b} \frac{\partial }{\partial b} +
             \hat{c} \frac{\partial }{\partial c},
\end{equation}
where a hat indicates a unit vector in the indicated direction.

\subsubsection{Perfectly correlated probability spaces}

We compare calculations when $x$ and $y$ are perfectly
correlated at points $(a,0,0,1-a)$ in the isomorphically
constrained space ${\cal P}_{\rm corr}$ and in the
non-constrained space ${\cal P}'_{\rm corr}$.

The joint entropy between $x$ and $y$ is
\begin{eqnarray}
 E_{xy}(a,b,c)
   &=& - a \log a - b \log b - c \log c \\
   && -  (1-a-b-c) \log(1-a-b-c) \nonumber
\end{eqnarray}
giving respective gradients in the ${\cal P}_{\rm corr}$ and
${\cal P}'_{\rm corr}$ spaces of
\begin{eqnarray}      \label{eq_entropy_max}
 \nabla E_{xy}|_{b=c=0}
   &=& - \hat{a} \log \left( \frac{a}{1-a} \right) \nonumber \\
 \nabla E_{xy}
   &=& - \hat{a} \log \left( \frac{a}{1-a-b-c} \right) \nonumber \\
    &&   - \hat{b} \log \left( \frac{b}{1-a-b-c} \right) \nonumber \\
    &&   - \hat{c} \log \left( \frac{c}{1-a-b-c} \right) \nonumber \\
 \lim_{(bc)\rightarrow (00)} \nabla E_{xy} &=& {\rm undefined}.
\end{eqnarray}
Equating these gradients to zero locates the maximum at
$(a,b,c)=(\frac{1}{2},0,0)$ in ${\cal P}_{\rm corr}$ and at
$(a,b,c)=(\frac{1}{4},\frac{1}{4},\frac{1}{4})$ in ${\cal
P}'_{\rm corr}$.

Writing $(a,b,c)=(p_1,p_2,p_3)$, the Fisher Information is a
matrix with elements $i,j\in\{1,2,3\}$ with
\begin{eqnarray}            \label{eq_Fisher_3d}
  F_{ij} &=&  \\
    && \hspace{-2cm} \sum_{xy} P_{xy}
     \left(\frac{\partial}{\partial p_i} \log P_{xy}\right)
      \left(\frac{\partial}{\partial p_j} \log P_{xy}\right). \nonumber
\end{eqnarray}
When isomorphically constrained in the space ${\cal P}_{\rm
corr}$, the Fisher Information is $F_{ij}|_{b=c=0}$ with the
only nonzero term being
\begin{eqnarray}                \label{eq_Fisher_Inf}
 F_{11} &=& (1-a) \left[ \hat{a} \frac{\partial }{\partial a} \log (1-a) \right]^2
     + a \left[ \hat{a} \frac{\partial }{\partial a} \log a \right]^2 \nonumber \\
     &=& \frac{1}{a(1-a)}
\end{eqnarray}
This means that the smaller the Variance the more the
information obtained about $a$. In the unconstrained space
${\cal P}'_{\rm corr}$, the Fisher Information is a very
different, $3\times 3$ matrix.

The likelihood function estimates probability parameters from
the observation of $n$ trials with $n_a$ appearances of event
$(x,y)=(0,0)$, $n_b$ appearances of event $(x,y)=(0,1)$, $n_c$
appearances of event $(x,y)=(1,0)$, and $n_d$ appearances of
event $(x,y)=(1,1)$.  We have $n_a+n_b+n_c+n_d=n$, giving the
Likelihood function
\begin{equation}
 L = f(n_a,n_b,n_c,n)  a^{n_a} b^{n_b} c^{n_c} (1-a-b-c)^{n-n_a-n_b-n_c}
\end{equation}
where $f(n_a,n_b,n_c,n)$ gives the number of combinations. The
optimization proceeds by evaluating the gradient of the Log
Likelihood function. When isomorphically constrained in the
space ${\cal P}_{\rm corr}$, the gradient of the Log Likelihood
function is
\begin{equation}
 \nabla \log L|_{b=c=0}
     =  \hat{a} \left[ \frac{n_a}{a} - \frac{n-n_a}{1-a}  \right],
\end{equation}
which equated to zero gives the optimal estimate at $a=n_a/n$
and $n_b=n_c=0$ as expected.  Conversely, when unconstrained in
the space ${\cal P}'_{\rm corr}$, the gradient of the Log
Likelihood function evaluates as
\begin{eqnarray}              \label{eq_max_likelihood_3d}
 \nabla \log L
     &=&  \hat{a} \left[ \frac{n_a}{a} - \frac{n-n_a-n_b-n_c}{1-a-b-c}  \right] \nonumber \\
     && + \hat{b} \left[ \frac{n_b}{b} - \frac{n-n_a-n_b-n_c}{1-a-b-c}  \right] \nonumber \\
     && + \hat{c} \left[ \frac{n_c}{c} - \frac{n-n_a-n_b-n_c}{1-a-b-c}  \right].
\end{eqnarray}
This is obviously a very different result, though at points
$(a,b,c)=(a,0,0)$ equating the log Likelihood to zero locates
the same estimate as before of $a=n_a/n$ and $n_b=n_c=0$.

In the unconstrained probability space ${\cal P}'_{\rm corr}$,
the expectation, variance, and entropy relations of interest
evaluate as
\begin{eqnarray}        \label{eq_expectation_relations_3d}
 \langle x\rangle - \langle y\rangle  &=& c-b  \nonumber  \\
 V(x) - V(y)   &=& (c-b)(a-d)    \\
 E_x &=& -\left[ (a+b) \log(a+b) + \right. \nonumber \\
 && \left. (1-a-b) \log(1-a-b) \right] \nonumber \\
 E_{xy}
   &=& -\left[ a \log a + b \log b + c \log c+ \right. \nonumber \\
   && \left. (1-a-b-c) \log(1-a-b-c)\right], \nonumber
\end{eqnarray}
which in the limit gives an undefined gradient
\begin{equation}      \label{eq_expectation_relations_3da}
 \lim_{(bc)\rightarrow (00)} \nabla \left[E_{xy}-E_x\right]
     = {\rm undefined}.
\end{equation}

\subsubsection{Independent probability spaces}

For the square die under consideration, we have probabilities
and expectations of
\begin{eqnarray}      \label{eq_ind_space}
  P_{xy}(00)-P_x(0) &=& ad - bc \nonumber \\
  \langle xy \rangle - \langle x \rangle \langle y \rangle &=&
      ad - bc \nonumber \\
  P_{x|y}(0|0)-P_x(0) &=& \frac{ad-bc}{a+c},
\end{eqnarray}
and entropies of
\begin{eqnarray}
  E_x &=& -(a+b) \log(a+b) - (1-a-b)\log(1-a-b) \nonumber \\
  E_y &=& -(a+c) \log(a+c) - (1-a-c)\log(1-a-c) \nonumber \\
  E_{xy} &=& -a \log a - b \log b - c \log c - d \log d,
\end{eqnarray}
giving gradients of
\begin{eqnarray}      \label{eq_ind_space2}
 \lim_{ad\rightarrow bc} \nabla \left[E_{xy}-E_x-E_y\right]
     &=&  \\
 && \hspace{-4.5cm}\lim_{ad\rightarrow bc}  \nabla \left\{
    a \log \left[\frac{d}{a}\frac{a-ad+bc}{d-ad+bc}\right] +
    b \log \left[\frac{d}{b}\frac{b+ad-bc}{d-ad+bc}\right] + \right.   \nonumber \\
 && \hspace{-4cm} \left. c \log \left[\frac{d}{c}\frac{c+ad-bc}{d-ad+bc}\right] +
    \log \left[\frac{d-ad+bc}{d}\right] \right\} \;\neq\; 0.\nonumber
\end{eqnarray}

\section{Optimizing simple decision trees}
\label{app_Optimizing_simple_decision_trees}

When the correlation (Eq. \ref{eq_rho_correlation}) between $x$
and $y$ is $\rho_{xy}=\rho$, and as long as both $p\neq 0$ and
$p\neq 1$, then the correlation constraint defines two surfaces
in the $(p,q,r)$ simplex at height
\begin{eqnarray}       \label{eq_r_plus_minus}
 r_{\pm}(p,q,\rho) &=&  \\
 && \hspace{-2.3cm} \frac{\rho^2 -2 q(1-p) (\rho^2 -1)
    \pm \rho  \sqrt{\rho^2+4q(1-q)\frac{(1-p)}{p}} }
    {2  \left[1 + p (\rho^2 -1) \right]}. \nonumber
\end{eqnarray}
The function $r_{+}(p,q,\rho)$ will give the required
correlation surfaces within the simplex. That is, when $\rho=0$
we have $r_+(p,q,0)=q$ as required.  Similarly, when $\rho=1$ we
have $r_+(p,q,1)\geq 1$ across the entire $(p,q)$ plane with the
equality $r_+(p,q,1)=1$ only where $q=0$ or $q=1$. We require
$\rho=1$ at $(q,r)=(0,1)$. Finally, when $\rho=-1$ and $x$ and
$y$ are perfectly anti-correlated, we have $r_+(p,q,-1)\leq 0$
across the entire $(p,q)$ plane with the equality
$r_+(p,q,-1)=0$ only where $q=0$ or $q=1$. We require $\rho=-1$
at $(q,r)=(1,0)$.

The strict requirement that $0\leq r_+(p,q,\rho)\leq 1$
establishes permissible regions on the $(p,q)$ plane. For
$0<\rho<1$, the permissible region is bounded by the $q=0$ line
and the line
\begin{equation}    \label{eq_permissible_region1}
  q(p,\rho) = \frac{p}{p+ \frac{\rho^2}{1-\rho^2}}.
\end{equation}
Similarly, for $-1<\rho<0$, the $(p,q)$ region is bounded by the
$q=1$ line and the line
\begin{equation}   \label{eq_permissible_region2}
  q(p,\rho) = \frac{1}{1+p \frac{1-\rho^2}{\rho^2}}.
\end{equation}

\end{document}